\documentstyle[prl,aps]{revtex}
\begin{document}


\vskip 4mm

\centerline{\large \bf Prismane C$_8$: a new form of carbon?}

\vskip 2mm

\centerline{Leonid A. Openov and Vladimir F. Elesin}

\vskip 2mm

\centerline{\it Moscow State Engineering Physics Institute
(Technical University)}
\centerline{\it 115409 Moscow, Russia}

\vskip 4mm

\begin{quotation}

Our numerical calculations on small carbon clusters point to the existence of
a metastable three-dimensional eight-atom cluster C$_8$ which has a shape of
a six-atom triangular prism with two excess atoms above and below its bases.
We gave this cluster the name "prismane". The binding energy of the prismane
equals to 5.1 eV/atom, i.e., is 0.45 eV/atom lower than the binding energy of
the stable one-dimensional eight-atom cluster and 2.3 eV/atom lower than the
binding energy of the bulk graphite or diamond. Molecular dynamics
simulations give evidence for a rather high stability of the prismane, the
activation energy for a prismane decay being about 0.8 eV. The prismane
lifetime increases rapidly as the temperature decreases indicating a
possibility of experimental observation of this cluster.

\end{quotation}

\vskip 4mm

Carbon is known to form a rich variety of crystal structures due to its ability
to exist in different valence states. As a result, coordination numbers in
carbon compounds range from two (e.g. carbyne), through three (e.g. graphite)
to four (e.g. diamond), while typical ("favorable") values of angles between
covalent bonds are 180$^o$ in carbyne, 120$^o$ within graphite layers, and
109$^o$28$^{\prime}$ in diamond. Recent discovery of the C$_{60}$ molecule
\cite{Kroto} and synthesis of C$_{60}$ clusters in macroscopic quantities
\cite{Kratschmer} have stimulated a renewed interest in carbon nanostructures
\cite{Lozovik}. A carbon nanostructure may be viewed as a graphite layer
transformed into a tube (nanotube), a ball (fulleren) {\it etc}, with an
exception that such nanostructures usually comprise pentagons along with
hexagons.

A discovery of the cubane C$_8$H$_8$ \cite{Eaton} appeared to be very
important from a fundamental viewpoint since the carbon atoms in the molecule
C$_8$H$_8$ are located in the apexes of a cube, so that the angles between
C-C bonds equal to 90$^o$, in contrast to the majority of carbon compounds.
As a consequence of "energetically unfavorable" bond angles, the molecule
C$_8$H$_8$ is metastable and accumulates a considerable amount of energy
\cite{Eaton}. Its cubic structure is stabilized by the corner hydrogen
atoms. This rises the question as to whether there exist stable or metastable
three-dimensional clusters that are composed of carbon atoms only and have
90$^o$ bond angles.

Another interesting question discussed in the literature
\cite{Weltner,Tomanek,Xu,Jones} is "What is the minimum number of atoms in
stable and metastable {\it three-dimensional} carbon clusters?". It must be
emphasized that while there is a substantial progress in studies of
relatively large carbon nanoparticles C$_N$ composed of $N\sim 100$ carbon
atoms \cite{Lozovik}, the controversy still remains concerning the structure
and energetics of small carbon clusters with $N\sim 10$, e.g., whether the
stable C$_4$, C$_6$, and C$_8$ clusters are cyclic or linear \cite{Weltner}.
Three-dimensional carbon clusters (cages) are believed to be stable for
$N\geq 20$ only \cite{Tomanek,Xu,Jones}, while for clusters with $N<20$ the
stable structures are one-dimensional linear chains or monocyclic rings
\cite{Weltner,Tomanek,Xu}. A characteristic feature of small carbon clusters
is the existence of metastable states whose binding energies are lower than
the binding energy of the stable cluster with the same $N$. For $N<14$, the
metastable structures are either one-dimensional (chains and rings) or
two-dimensional (graphite flakes), see, e.g., Ref. \cite{Tomanek}. To our
knowledge, up to now there was no experimental and theoretical evidence for
neither stable nor metastable {\it three-dimensional} carbon clusters with
$N<14$ (the occurrence of metastable cage C$_{14}$ has been reported in
Ref. \cite{Jones} based on the results of density functional calculations).

The purpose of this work was to search for metastable three-dimensional
clusters C$_N$ with $N<14$ by means of numerical simulation. One of such
clusters, the cluster C$_8$, was found. Here we report the numerically
calculated structural and energetical characteristics of this cluster whose
lifetime appears to be surprisingly long due to rather high value of its
activation energy, about 0.8 eV.

We have carried out TBMD (tight-binding molecular dynamics) simulations of
small carbon clusters making use of a transferable tight-binding potential
recently developed for carbon by Xu {\it et al.} \cite{Xu,Xu2}. This
numerical technique allows one to calculate the total energy of the cluster
having an arbitrary atomic configuration. It had been proven to reproduce
accurately the energy-versus-volume diagram of carbon polytypes and to give a
good description of both small clusters and bulk structures of carbon
\cite{Xu,Xu2}. We have checked that this technique describes the structure
and energetics of small carbon clusters quite well, the difference in bond
lengths and binding energies between our results and available
{\it ab initio} calculations \cite{Weltner} usually did not exceed 10$\%$.

The binding (cohesive) energy $E_{coh}(N)$ of $N$-atom cluster C$_N$ has been
determined as (see, e.g., \cite{Tomanek})
\begin{equation}
E_{coh}(N)=NE(1)-E(N),
\label{Ecoh}
\end{equation}
where $E(N)$ is the total energy of the cluster, $E(1)$ is the energy of an
isolated carbon atom. The positive value of $E_{coh}$ points to the stability
of the cluster with respect to its fragmentation into $N$ carbon atoms. At a
given $N$, there may exist several atomic configurations having $E_{coh}>0$.
The cluster with the highest value of $E_{coh}$ is stable while the clusters
with lower (but positive) values of $E_{coh}$ are metastable. A metastable
cluster can transform to the stable, energetically favorable configuration. A
characteristic time of such a process (the lifetime $\tau$) depends on the
height of the energy barrier separating the metastable and stable
configurations.

We have thoroughly analyzed a number of atomic configurations as possible
candidates to metastable three-dimensional carbon structures. However, all of
them (with one exception that constitutes the essence of this paper, see
below) appeared to be unstable and transformed to one-dimensional clusters or
decayed into small fragments even at $T\rightarrow 0$. In particular, we have
checked for a possibility of existence of a three-dimensional cluster C$_8$
having the cubic structure (by analogy with the cubane C$_8$H$_8$
\cite{Eaton}), but we have found that such a cluster is unstable.
Nevertheless, we have discovered that the eight-atom cluster C$_8$ can exist
as a three-dimensional cluster having a shape of a six-atom triangular prism
with two excess top atoms above and below its bases. We gave this cluster the
name "prismane". It is shown in Fig. 1.

The binding energy of the prismane equals to $E_{coh}(8)/8=5.1$ eV/atom. This
is 0.45 eV/atom lower than the binding energy of the stable one-dimensional
eight-atom cluster and 2.3 eV/atom lower than the binding energy of the bulk
graphite or diamond. Hence, the prismane C$_8$ is metastable. The lengths of
C-C bonds equal to 2.31 {\AA} within each base, 1.28 {\AA} between the two
bases, and 1.47 {\AA} between each base and the nearest top atom, see Fig. 1.
We note that the angles between C-C bonds within the bases and C-C bonds
connecting the bases are equal to 90$^o$. This is one reason for a lower value of the binding energy of
the prismane relative to the binding energy of the stable one-dimensional
eight-atom cluster. Meanwhile, the angles between C-C bonds connecting the
top atoms with bases of the prism equal to 104$^o$, being close to the value
of bond angle in bulk diamond (109$^o$28$^{\prime}$).

In order to determine the energy barrier separating the metastable prismane
structure and the stable chain structure, we have carried out the molecular
dynamics simulations of the prismane decay for different values of the
initial temperature $T_{ini}$. The time of one molecular dynamics step
constituted $t_0 = 2.72 \cdot 10^{-16}$ s, about one percent of oscillation
period of the dimer C$_2$. The temperature $T$ of the cluster was calculated
after each 500 molecular dynamics steps according to the formula
\begin{equation}
\frac{3}{2}k_B T = \langle E_{kin} \rangle,
\label{temperature}
\end{equation}
where $k_B$ is the Boltzmann constant, $\langle E_{kin} \rangle$ is the
kinetic energy per atom averaged over a period of time $\Delta t=500t_0$.
Such a time averaging enables one to avoid strong fluctuations of $T$
stemming from the small number of atoms in the cluster under investigation.
The initial temperature $T_{ini}$ was determined from Eq. (\ref{temperature})
at $0\le t \le 500t_0$.

On general grounds, one would expect the probability of the cluster decay in
a unit of time, $W$, to be given by a statistical formula
\begin{equation}
W=W_0 \exp(-E_a/k_B T_{ini}),
\label{probability}
\end{equation}
where the factor $W_0$ has dimensionality s$^{-1}$, and $E_a$ is the
activation energy, i.e., the height of the energy barrier separating the
given metastable state from the stable or some other metastable state of the
cluster. The cluster lifetime $\tau$ may be defined as
\begin{equation}
\tau=1/W=\tau_0 \exp(E_a/k_B T_{ini}),
\label{lifetime}
\end{equation}
where $\tau_0=1/W_0$. Molecular dynamics simulations allow one to determine
the lifetime $\tau$ directly at any particular value of $T_{ini}$ as the time
of cluster decay into another atomic configuration. It is convenient to go
from the cluster lifetime $\tau$ to a critical number of molecular dynamics
steps $N_c=\tau/t_0$ which it takes for the cluster to decay:
\begin{equation}
N_c=N_0 \exp(E_a/k_B T_{ini}),
\label{N_c}
\end{equation}
where $N_0=\tau_0/t_0$.

Figure 2 shows a typical example of the binding energy per atom,
$E_{coh}(8)/8$, and the temperature $T$ of the prismane as functions of the
number of molecular dynamics steps $N$ for the case $T_{ini}=950$ K. One can
see that $E_{coh}(8)/8$ and $T$ fluctuate near the values 5.0 eV/atom and
950 K respectively over $N\approx 23500$ molecular dynamics steps. The
prismane preserves its form during this period of time. At $N_c=23600\pm100$
the prismane decays into the stable eight-atom chain. The temperature rises
up to 2600 K. Strong oscillation of the atoms within the chain with respect
to one another brings about the variation of $E_{coh}(8)/8$ in time from
4.9 eV/atom up to the binding energy of equilibrium chain, 5.55 eV/atom.

Note however that the lifetime $\tau=N_c t_0=6.4\cdot 10^{-12}$ s is rather
short on a macroscopic scale, while we are interested in phenomena which
happen on time scales of seconds or even months and years. According to
Eq. (\ref{lifetime}), such long lifetimes are expected for temperatures much
lower than the value $T_{ini}=950$ K used in simulations shown in Fig. 2. But
those lifetimes are impossible to achieve by means of direct computer
simulations.

To overcome the problem, we have calculated numerically the critical values
of $N_c$ for several values of $T_{ini}$ and fitted the numerical data by
Eq. (\ref{N_c}). The results are shown in Fig. 3. We stress that the process
of cluster decay is probabilistic in nature. Hence, the lifetime $\tau$ is
not uniquely determined at a given value of $T_{ini}$. As a result, the data
for $\ln (N_c)$ as a function of $1/T_{ini}$ are somewhat scattered, see
Fig. 3. Nevertheless one can see that the numerical data can be well fitted
by a straight line over rather wide ranges of $N_c=300\div 206000$ and
$T_{ini}=890\div 1700$ K values, in accordance with Eq. (\ref{N_c}). The
coefficients of this fit give the values of $N_0=2.1$ and $E_a/k_B=9500$ K
in Eq. (\ref{N_c}).

We note that the activation energy $E_a=0.82$ eV is very large, thus
resulting in relatively high stability of the prismane. Indeed, since the
dependence of $\tau$ on $T_{ini}$ is exponentially strong, a decrease in
$T_{ini}$ leads to a rapid increase in $\tau$, so that $\tau \sim 10\mu s$ at
$T_{ini}=400$ K. An important remark is in order here. Since the total energy
of the cluster is conserved in our simulations, the cluster is unable to
decay if the {\it maximum} attainable (at a given value of $T_{ini}$) kinetic
energy $E_{kin}^{max}$ of the cluster is less than $E_a$. In its turn,
$E_{kin}^{max}$ is twice the {\it time-averaged} kinetic energy at the
initial stage of cluster evolution, i.e.,
$E_{kin}^{max}=2\cdot 8 \cdot \frac{3}{2}k_B T_{ini} = 24k_B T_{ini}$, see
Eq. (\ref{temperature}). Hence, at $T_{ini}<E_a/24k_B\approx 400$K the
lifetime of the cluster equals to infinity. Thus, in general, an
extrapolation of $\tau(T_{ini})$ curve to the range $T_{ini}<400$K
is incorrect.

The lifetime $\tau$ is extremely sensitive to the value of the activation
energy $E_a$ since the latter appears in the exponent, see
Eq. (\ref{lifetime}). Hence, even a minor change in $E_a$ will result in a
substantial variation of $\tau$ at a given $T_{ini}$. In order to refine the
value of $E_a$ it is necessary to calculate the critical values of $N_c$ for
a greater number of initial temperatures $T_{ini}$. However, we believe it is
unlikely that such a refinement will cause the value of $E_a$ to change
significantly.

Finally, it should be stressed that we have confirmed the existence of the
three-dimensional metastable prismane structure by making use of other
computer codes, e.g., MOPAC and that based on the empirical interatomic
potential proposed for carbon systems by Tersoff \cite{Tersoff}. The overall
shape of the cluster (Fig. 1) appeared to be the same, while the bond lengths
and the binding energy were to some extent different. So, the prismane
C$_8$ certainly is not just an artifact of a specific simulation technique
used in this study, while the values of the prismane binding energy,
activation energy, and lifetimes at different temperatures may be refined by
means of more sophisticated calculations.

In conclusion, we have predicted the existence of a three-dimensional cluster
C$_8$ (prismane) which is the smallest three-dimensional carbon cluster found
so far experimentally or theoretically. This cluster appears to be metastable
and has the binding energy 0.45 eV/atom below the binding energy of the
stable one-dimensional eight-atom cluster. However, molecular dynamics
simulations point to a rather high stability of the prismane suggesting that
this cluster may be observed experimentally.

We are grateful to A. V. Krasheninnikov and N. E. L'vov for the help in
preparation of the manuscript. The work was supported in part by the Russian
State Program "Integration" and by the Contract DSWA01-98-C-0001.

\newpage
\centerline{\bf Figure captions}
\vskip 2mm

Fig. 1. Prismane C$_8$. Bond lengths: $d_{AB}=2.31$ {\AA},
$d_{AC}=1.28$ {\AA}, $d_{AD}=1.47$ {\AA}. Bond angles: $\angle BAC = 90^o$,
$\angle ADB = 104^o$, $\angle ABE = 60^o$

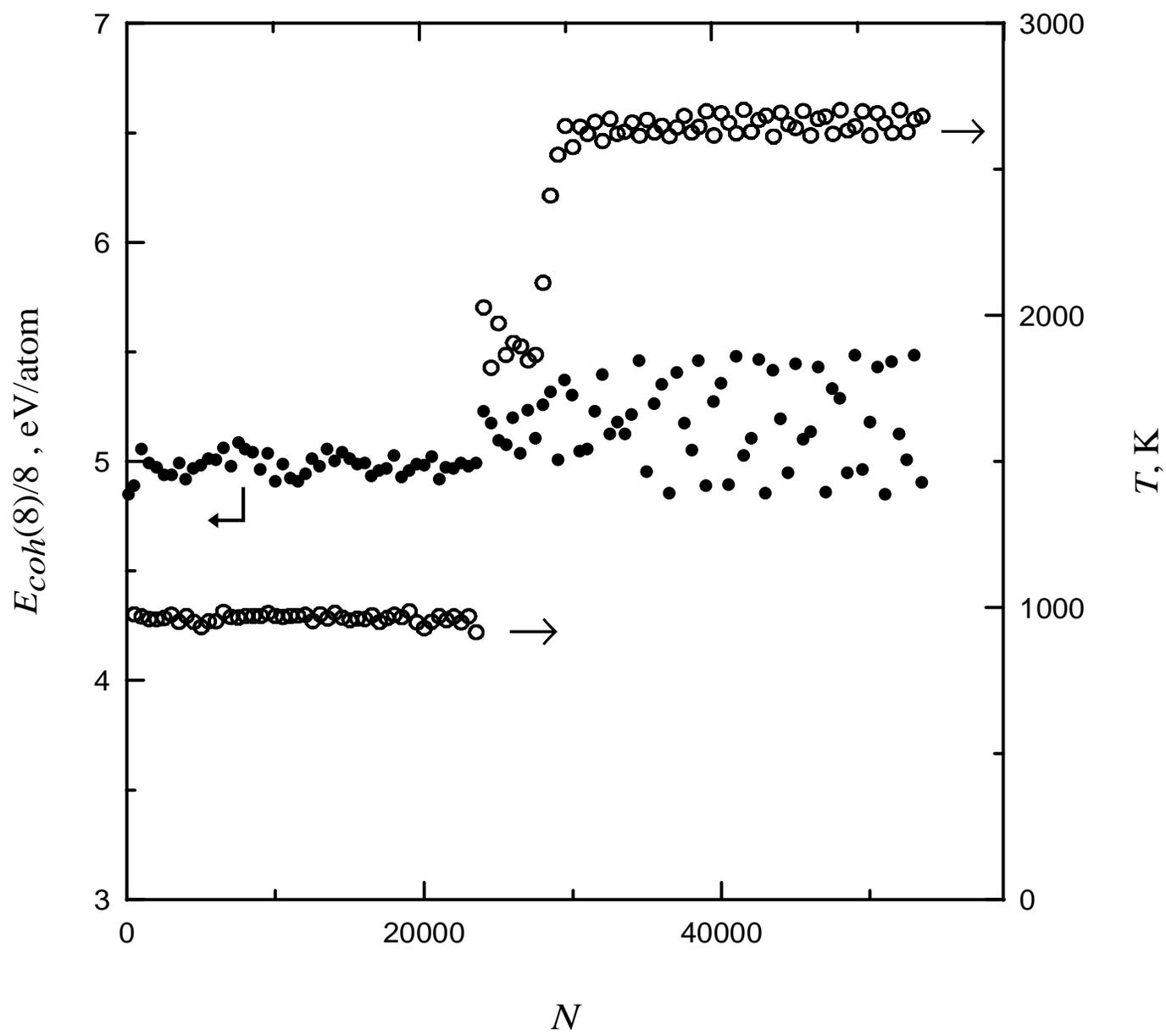
Fig. 2. Cohesive energy per atom $E_{coh}(8)/8$ (closed circles) and
temperature $T$ (open circles) of the prismane C$_8$ versus the number of
molecular dynamics steps $N$. The time of one step
$t_0 = 2.72 \cdot 10^{-16}$ s. Initial temperature $T_{ini}=950$ K. The
prismane decays at $N_c=23500\div 23700$. The lifetime
$\tau=N_c t_0=6.4\cdot 10^{-12}$ s.

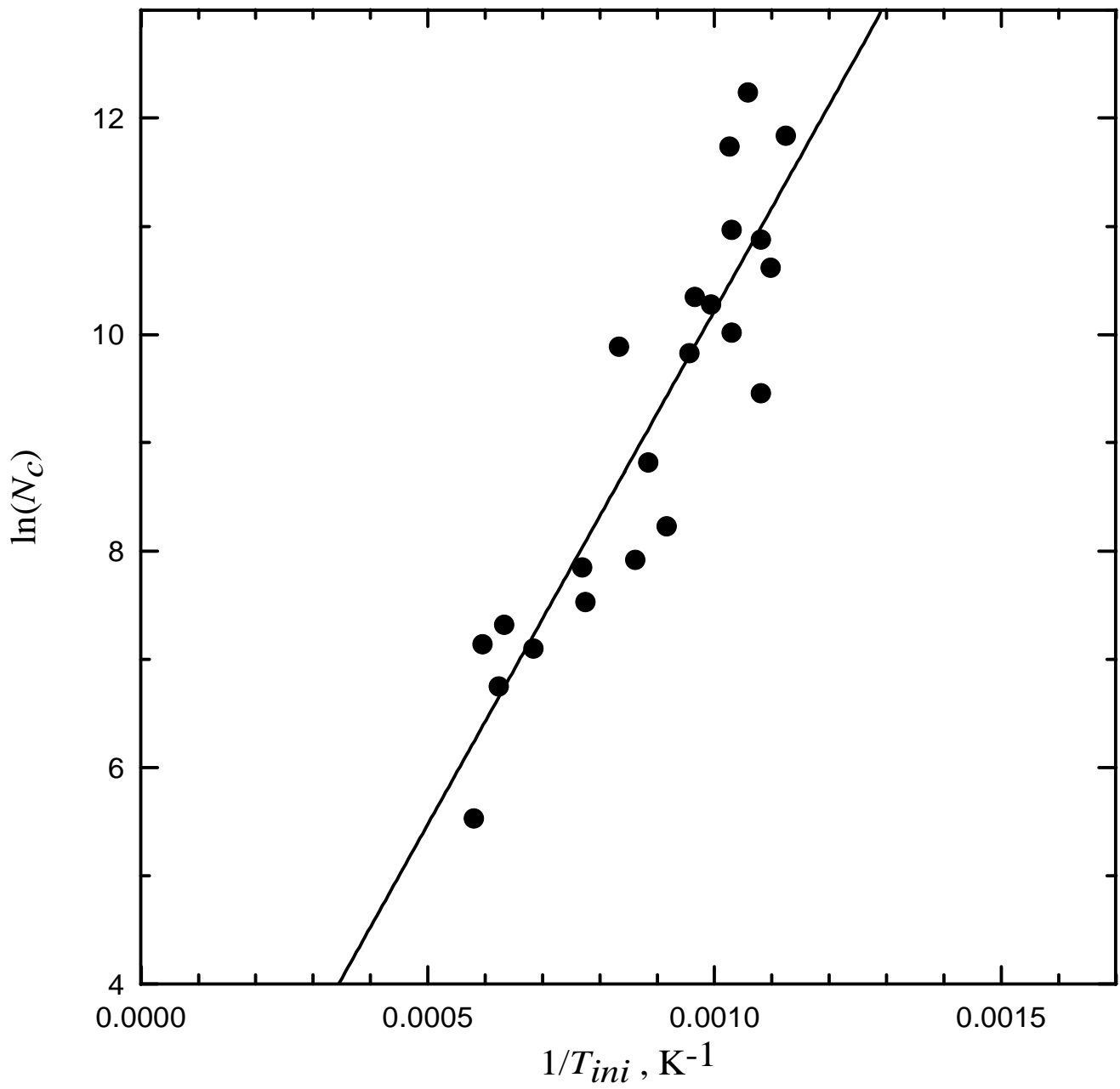
Fig. 3. Plot of the logarithm of the critical number of molecular dynamics
steps $N_c$ (corresponding to the prismane decay) versus the inverse initial
temperature $1/T_{ini}$. Circles are the results of numerical calculations.
The solid line is the least-square fit $\ln (N_c)=0.7263+9496/T_{ini}$,
where $T_{ini}$ is measured in K.


\vskip 4mm

\begin{references}

\bibitem{Kroto} H. W. Kroto {\it et al.}, Nature {\bf 318}, 162 (1985).
\bibitem{Kratschmer} W. Kr\"atschmer {\it et al.}, Nature {\bf 347}, 354
(1990).
\bibitem{Lozovik} Yu. E. Lozovik and A. M. Popov, Usp. Fiz. Nauk {\bf 167},
751 (1997).
\bibitem{Eaton} P. E. Eaton and G. J. Castaldi, J. Am. Chem. Soc.
{\bf 107}, 784 (1985), and references therein.
\bibitem{Weltner} W. Weltner, Jr., and R. J. Van Zee, Chem. Rev. {\bf 89},
1713 (1989).
\bibitem{Tomanek} D. Tom\'anek and M. A. Schluter, Phys. Rev. Lett. {\bf 67},
2331 (1991).
\bibitem{Xu} C. H. Xu {\it et al.}, Phys. Rev. B {\bf 47}, 9878 (1993).
\bibitem{Jones} R. O. Jones and G. Seifert,
Phys. Rev. Lett. {\bf 79}, 443 (1997).
\bibitem{Xu2} C. H. Xu {\it et al.},
J. Phys.: Condens. Matter {\bf 4}, 6047 (1992).
\bibitem{Tersoff} J. Tersoff, Phys. Rev. Lett. {\bf 61}, 2879 (1988).

\end{references}
\end{document}